\documentclass[
 amsmath,
 amssymb,twocolumn,superscriptaddress,
 aps,prx]{revtex4-2}

\preprint{APS/123-QED}

\usepackage{graphicx}
\usepackage{dcolumn}
\usepackage{hyperref}
\usepackage{float}
\usepackage{braket}
\usepackage[normalem]{ulem}
\usepackage{soul}

\begin{document}
\preprint{APS/123-QED}

\title{Quantum machine learning  via continuous-variable cluster states and teleportation}

\author{Jorge Garc{\'\i}a-Beni}
\affiliation{
 Instituto de F{\'\i}sica Interdisciplinar y Sistemas Complejos (IFISC), UIB–CSIC \\
 UIB Campus, Palma de Mallorca, E-07122, Spain}
\author{Iris Paparelle}
\affiliation{Laboratoire Kastler Brossel, Sorbonne Universit\'e, ENS-Universit\'e PSL, CNRS, Coll\`ege de France, 4 place Jussieu, 75252 Paris, France}
\author{Valentina Parigi}
\affiliation{Laboratoire Kastler Brossel, Sorbonne Universit\'e, ENS-Universit\'e PSL, CNRS, Coll\`ege de France, 4 place Jussieu, 75252 Paris, France}
\author{Gian Luca Giorgi}
\affiliation{
 Instituto de F{\'\i}sica Interdisciplinar y Sistemas Complejos (IFISC), UIB–CSIC \\
 UIB Campus, Palma de Mallorca, E-07122, Spain}
\author{Miguel C. Soriano}
\affiliation{
 Instituto de F{\'\i}sica Interdisciplinar y Sistemas Complejos (IFISC), UIB–CSIC \\
 UIB Campus, Palma de Mallorca, E-07122, Spain}
\author{Roberta Zambrini}
\affiliation{
 Instituto de F{\'\i}sica Interdisciplinar y Sistemas Complejos (IFISC), UIB–CSIC \\
 UIB Campus, Palma de Mallorca, E-07122, Spain}
\email{roberta@ifisc.uib-csic.es}

\date{\today}
\begin{abstract}

A new approach suitable for distributed quantum machine learning and exhibiting memory is proposed for a photonic platform. This measurement-based quantum reservoir computing takes advantage of continuous variable cluster states as the main quantum resource. Cluster states are key to several photonic quantum technologies, enabling universal quantum computing as well as quantum communication protocols. The proposed measurement-based quantum reservoir computing is based on a neural network of cluster states and local operations, where input data are encoded through measurement, thanks to quantum teleportation. In this design, measurements enable input injections, information processing and continuous monitoring for time series processing. The architecture's power and versatility are tested by performing a set of benchmark tasks showing that the protocol displays internal memory and is suitable for both static and temporal information processing without hardware modifications. This design opens the way to distributed machine learning.

\end{abstract}

\maketitle

\section{Introduction} \label{intro-sec}
The use of quantum substrates for machine learning has been motivated by the quest for a quantum advantage as well as the availability of NISQ devices \cite{Biamonte2017,dunjko2018machine,havlivcek2019supervised,Abbas2021,Cerezo2021}. 
The potential and challenges of variational circuits and quantum kernels have been established mainly in superconducting circuits 
 \cite{Cerezo2022}.
With their inherent scalability, robustness, and ability to operate at room temperature, photonic quantum technologies, provide a promising alternative for addressing the growing demand for fast and efficient processing of both classical and quantum data \cite{photQtech2023}. Quantum photonic platforms have already proven remarkable achievements in several fields, most notably in communications but more recently also in computation \cite{boson-sampling-1,boson-sampling-2,boson-sampling-3,Peruzzo2014,ising-machines-1,ising-machines-2}.
Gaussian boson sampling experiments have exhibited quantum advantage, achieving unprecedented quantum speedup compared to current classical supercomputers \cite{boson-sampling-1,boson-sampling-2,boson-sampling-3}. Variational eigensolvers executed on photonic quantum processors \cite{Peruzzo2014} have demonstrated a more efficient utilization of quantum resources compared to other algorithms, such as quantum phase estimation. Moreover, quantum states of light have been implemented in Ising Machines, proving more effective than current algorithms in specific scenarios \cite{ising-machines-1,ising-machines-2}.

The pursuit of fault-tolerant quantum computing in photonic platforms \cite{Bourassa2021} shares some resources with Gaussian boson sampling setups and builds on the so-called measurement-based quantum computing (MBQC) approach \cite{Briegel2001, Raussendorf2002}. 
In this framework, cluster states serve as a crucial resource. These highly entangled states enable universal quantum computations through the application of local operations. MBQC is currently pursued as a promising approach in quantum photonics, initially formulated with discrete variables (DVs) encoding\cite{Nielsen2004,Browne2005}, and later with continuous variables (CV) encoding\cite{Zhang2006,Menicucci2006,Gu2009}. DV cluster states have been experimentally realized \cite{Walther2005,Kiesel2005,Bell_2013}. However, despite recent progress \cite{Schwartz22,Thomas22} deterministic scalability has not been reached yet. Conversely, continuous variable (CV) quantum cluster states can be generated deterministically \cite{Zhang2006}. Leveraging temporal multiplexing techniques \cite{Menicucci2010,Menicucci2011}, experimental clusters have achieved unprecedented  sizes \cite{furusawa_big,furusawa_2d_cluster}. Multi-mode quantum gates have also been successfully implemented on these temporal cluster platforms \cite{cluster_gates_1,cluster_gates_2}. Furthermore, frequency multiplexing has demonstrated promising experimental outcomes both in scale \cite{60_frequency_cluster} and reconfigurability \cite{Cai2017}. Optical cluster states have applications beyond computation, including quantum communications \cite{Azuma2015,Cai2017} and error correction protocols \cite{Bell2014}. They are also being explored as a promising resource for blind and quantum distributed computing protocols and the construction of the quantum internet \cite{Barz12,Koudia_2024}.

Within the context of photonics, quantum optical neural network architectures have also been proposed as promising candidates for quantum machine learning (QML), ranging from regular feedforward neural networks in DVs \cite{optical_nn} and CVs \cite{cv_optical_nn} to convolutional \cite{qu_CNNs} and recurrent neural network \cite{recurrent_optical_nn} designs. In this work, we propose a novel design that extends the application of CV clusters to encompass QML tasks. Our platform operates in the quantum reservoir computing (QRC) paradigm, which is a less demanding QML framework that is not restricted to circuit design and requires fewer optimization resources \cite{opportunities}. Reservoir computing \cite{RCbook,tanaka} is a machine learning framework for processing time-dependent data using a fixed, high-dimensional dynamical system. It features an internal memory allowing for online processing of data series, as recurrent neural networks, but with the advantage of efficient training and reduced computational cost in multi-task operations \cite{jaeger2004harnessing,Appeltant2011,Torrejon2017,abreu2024photonics}. Following successful implementation in photonic systems \cite{brunner2013parallel,Vandoorne2014,Larger,PhotRC_book,rafayelyan2020large,lupo2023deep},
optical platforms have been recently proposed also in quantum substrates, including integrated circuits using bosonic networks  \cite{Nokkala2021,Dudas2023,Guillem2023},  multimode light networks in feedback loops \cite{GBeni2023}, quantum memristors \cite{memristor}, and a recent experimental implementation in superconducting qubit-cavity architecture \cite{analogue_QRC_experimental}. 

In QRC input is injected via external driving and processed on a quantum substrate and measured observables constitute the output layer \cite{fujii-nakajima}. Continuous monitoring poses a challenge to online processing in quantum settings \cite{mujal_meas} and photonic solutions based on beam splitters and homodyne detection have been proposed \cite{GBeni2023}. In this work, we propose a CV optical QRC platform that utilizes cluster states as resources and where, beyond continuous monitoring, measurement also enables input injection and dynamical driving of the reservoir. Performance in static and temporal tasks as well as internal memory are analyzed. This architecture deviates from previous classical and quantum designs establishing cluster states as a resource also in machine learning, opening the ways to distributed machine learning.

\section{Measurement based quantum reservoir computing} \label{platform-sec}
\begin{figure}[t]
    \centering
    \includegraphics[width=\linewidth]{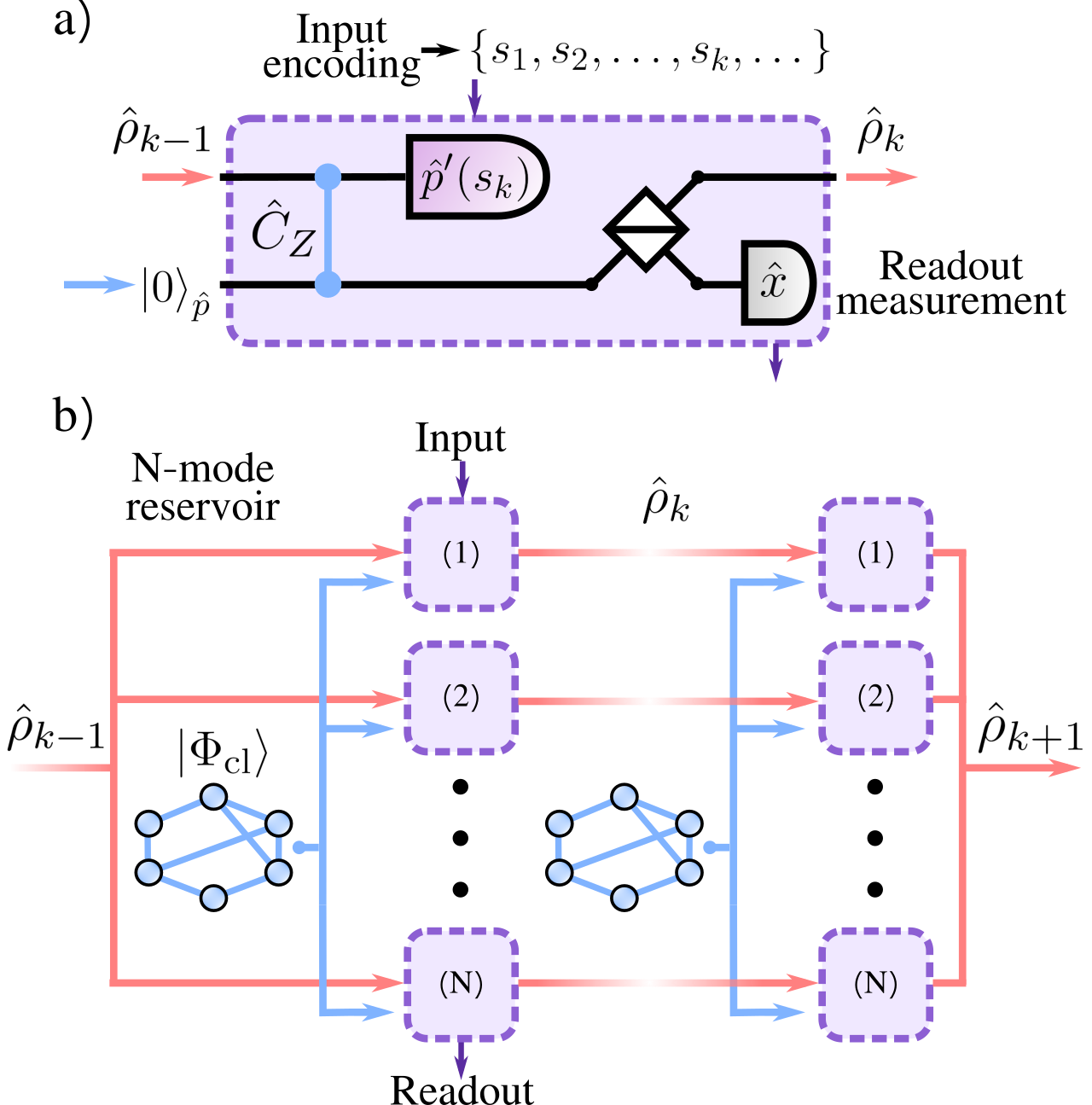}
    \caption{Reservoir computer scheme representation of: (a) single mode QRC protocol, using the reservoir state, $\hat{\rho}_{k-1}$, as an auxiliary state, a $\ket{0}_{\hat{p}}$ eigenstate and an ideal $\hat{C}_{Z}$ gate for teleportation. The input, $s_{k}$, is taken from a sequence and encoded in the measurement basis of the teleportation protocol. After that, a BS coupled to the vacuum and a homodyne detector are placed to monitor the system. The whole protocol is repeated for the quantum state that comes out, $\hat{\rho}_{k}$, using the next input in the sequence. The entire process can be encompassed in a cell (purple dashed box) comprising local operations; (b) the multimode QRC protocol now employs a cluster state as a resource. Each single reservoir mode is coupled to one cluster mode and both undergo the sequence of operations from the cell of Fig. \ref{figure-1}a. Using a cluster state enables applying a global input-dependent basis using local operations (inside each cell) at every time step.}
    \label{figure-1}
\end{figure}

We start by recalling the basic properties of quantum teleportation in CV systems in Sec. \ref{sec:recall}. 
We then build on this prior knowledge and describe our suggestion for a quantum measurement-based reservoir computing platform in Sec. \ref{sec:platform}.
In particular, our suggested platform is illustrated in Fig. \ref{figure-1}.
\subsection{Recall of quantum teleportation in CVs\label{sec:recall}}
Teleportation is key for quantum information processing in MBQC using cluster states \cite{Gu2009}. A basic CV teleportation circuit to teleport an arbitrary one-mode auxiliary state, $\ket{\psi_{\text{aux}}}$, uses a $\hat{p}$-quadrature eigenstate, $\ket{0}_{\hat{p}}$, as a resource. These states are entangled using a controlled-$Z$ ($\hat{C}_{Z}$) gate, $\hat{C}^{(A,B)}_{Z}(\xi) = \exp\left[-i\xi \hat{x}_{A} \otimes \hat{x}_{B}\right]$. The parameter $\xi$ determines the strength of the $\hat{C}_{Z}$ gate and ideal teleportation, which is the one considered in this work, is achieved when $\xi = 1$. Then, the observable $\hat{p}_{i}'(s) = \hat{p}_{i} + s \hat{x}_{i}$ is measured on the auxiliary mode, enabling the teleportation on the remaining mode, which becomes $\ket{\psi_{\text{QTP}}} \propto e^{-i s \hat{x}^{2}} \ket{\psi_{\text{aux}}}$. So when the parameter $s = 0$ a simple teleportation is performed, while when $s \neq 0$ the state is teleported with an added quadratic phase gate, function of the parameter $s$. This depends only on the measured observable $\hat{p}_{i}'(s)$, which can be experimentally accessed via phase rotation and field amplification \cite{Miwa2009}. If we have an $N$-mode auxiliary state, $\ket{\Psi_{\text{aux}}}$, multimode gates can be implemented by using an $N$-mode cluster state as a resource, 
\begin{equation} \label{cluster-eq}
\ket{\Phi_{\text{cl}}} = \prod_{(i,j) \in \mathcal{G}} \hat{C}^{(i,j)}_{Z} (\xi_{ij}) \bigotimes_{k=1}^{N} \ket{0}_{\hat{p}_{k}} \ ,
\end{equation}
where $\mathcal{G}$ stands for the graph topology of the cluster. We have considered the cluster $\hat{C}_{Z}$ gate's strengths $\xi_{ij} = 1$ (ideal gates) unless stated otherwise. If each mode on the auxiliary state, $\ket{\Psi_{\text{aux}}}$, and each mode of the cluster state, $\ket{\Phi_{\text{cl}}}$, are coupled one-to-one via ideal $\hat{C}_{Z}$ gates and measured in mode-dependent basis $\hat{p}'_{i}(s_{i})$, $i = 1, 2, \dots, N$ (i.e. each mode goes through a local basic teleportation circuit), then the remaining state becomes $\propto \hat{U}(\mathbf{s}) \ket{\Psi_{\text{aux}}}$ where $\hat{U}(\mathbf{s})$ is a unitary. 
The teleportation state has a multi-mode entangling gate applied to it (See App. Sect. \ref{app-QTP}). This multimode gate is a function of the chosen parameter vector $\mathbf{s} = \left( s_{1}, s_{2}, \dots, s_{N} \right)$, where each value of the vector has been encoded in the measurement basis of each mode. The resource cluster state allows global entangling gates to be implemented using local operations. This teleportation protocol is an idealization, as $\ket{0}_{\hat{p}}$ states are not physical, requiring an infinite amount of squeezing to be produced. Using finitely squeezed $p$-quadrature states instead allows teleportation with an added noise term that decays exponentially with the amount of squeezing \cite{Gu2009}. Current experiments have reached squeezing levels high enough for the non-ideal cluster states to be non-separable \cite{furusawa_2d_cluster,60_frequency_cluster} and achieved high fidelity for the implemented gates with teleportation \cite{cluster_gates_1}.
\subsection{Quantum reservoir computing platform\label{sec:platform}}
QRC protocols are designed to process sequential data and for each data of the time series ($\{\mathbf{s}_k\}_k$) usually information is processed in three layers: (i) encoding the external input into the quantum substrate (input layer); (ii) letting the quantum reservoir evolve in time according to a certain Hamiltonian or master equation evolution (reservoir layer) and (iii) measuring a set of observables of the quantum system to approximate a target task (readout layer where linear regression is realized). The protocol is then repeated for the following input in the sequence (more details on reservoir computing in App. Sect. \ref{app-RC}). 
Typically, the input is encoded (i) through state reset  \cite{fujii-nakajima}, or parametrized quantum circuit gates  \cite{Chen2019,Chen2020}, or by applying a controlled driving into the Hamiltonian  \cite{Govia2021,Sannia2024,yelin}. Here, we pursue an alternative strategy and take advantage of the tunability of the teleportation protocol to encode the external input $\mathbf{s}_k$ on the measurement basis. 

As illustrated in Fig. \ref{figure-1}a, at each $k$-th time step, the quantum reservoir state, represented by the density matrix $\hat{\rho}_{k-1}$, undergoes teleportation. 
The input dependent quadrature, $\hat{p}'(s_{k})$ (where for simplicity we consider a scalar input) is measured so an input dependent gate acts on the teleported (and new reservoir) state. The state of the reservoir needs to be sequentially monitored in order to extract information at each time step. Hence, after teleportation, the reservoir mode passes through a beam-splitter (BS) of transmissivity $T$, interacting with vacuum, depicted as a diamond in Fig. \ref{figure-1}a. One of the output modes is measured via homodyne detection of the $x$-quadrature, while the other is the (monitored) reservoir. Before seeing how to scale this design beyond the single mode, we highlight some key features to gain physical insight into the measurement based QRC. This way of encoding the input via measurement represents a change of perspective with respect to the three layer reservoir computing as both input encoding (i) and reservoir evolution (ii) are locally realized via measurement and teleportation.
As for the following pass through the BS and the homodyne detection, this realizes two essential functions in QRC. First, it allows us to monitor the quantum system at each time step, providing the readout layer (step (iii)) while maintaining a quantum reservoir state with memory (the unmeasured path $\hat{\rho}_{k}$ in Fig. \ref{figure-1}a). Second, the losses added passing through the BS induce a non-unitary dynamics of the quantum reservoir state leading to \textit{fading memory}. This property is the ability of the output variables to depend only on the recent history of the input sequence and is a necessary condition for QRC \cite{GRIGORYEVA2018495}. 

 The extension of the protocol to accommodate high dimensional reservoirs is done by exploiting the multimode teleportation revisited above in Sec. \ref{sec:recall}. Now, at time step $k$, our reservoir comprises an $N$-mode quantum state with density matrix $\hat{\rho}_{k-1}$ and an $N$-mode cluster state, $\hat{\Phi_{\text{cl}}}$, is used as a resource. The cluster is depicted as a network in Fig. \ref{figure-1}b.  Each mode in the reservoir is coupled one-to-one to the corresponding mode in the cluster, undergoing the same sequence of local operations as in the single mode case, depicted as purple-dashed ``cells'' in Fig. \ref{figure-1}b.
Hence, in each local cell (consisting in the circuit from Fig. \ref{figure-1}a) the input $s_{k}$ is injected through the measurement basis choice
\begin{equation} \label{p-prime-sk-eq}
    \hat{p}_{i}'(s_{k}) = \hat{p}_{i} + \left( \alpha_{i} s_{k} + \beta_{i} \right) \hat{x}_{i} \ \ (i = 1, 2, \dots N) \ ,
\end{equation}
where $\alpha_{i}$ and $\beta_{i}$ are mode-dependent parameters that are tuned depending on the task. As we measure the $x$-quadrature of each reservoir mode, the observables that are taken for the readout layer are the second-order moments, $\left\{ \left\langle \hat{x}_{i}^{(k)} \hat{x}_{j}^{(k)} \right\rangle \right\}_{i,j = 1}^{N} \equiv \left\{ O_{l}^{(k)} \right\}_{l=1}^{N(N+1)/2}$, including correlations among modes. This choice of observables enables non-linearities in the readout layers, as the implemented gates are linear in the field quadratures (see App. Sect. \ref{app-dynamics}). In that regard, measuring second-order moments suits the same function as measuring the field intensity in other classical photonic reservoir computing platforms \cite{PhotRC_book}. Furthermore, as  {the protocol we are proposing implements} quadratic gates, the resulting quantum states  {will} remain Gaussian \cite{Gu2009}. Gaussian operations have been shown to achieve universality in QRC, as described in Ref. \cite{Nokkala2021}. This implies the ability to approximate any time-invariant, causal, and fading memory map arbitrarily well using elements from our QRC class. Importantly, this notion of universality differs from that in quantum computing. For instance, cubic gates like  $e^{-i s \hat{x}^{3}}$ are essential for universal MBQC.

Finally, the readout function is built by performing a regression on the readout observables. For example, in a linear regression $y_{k} = w_{0} + \sum_{l=1}^{N(N+1)/2} w_{l} O_{l}^{(k)}$ the set of weights, $w_{l}$, is optimized so the predicted function converges to a target sequential function $\bar{y}_{k}$, which is trained to be reproduced by the reservoir. The resulting state at time step $k$ after the local operations have been performed can be written as a mixed state, $\hat{\rho}_{k} = T \hat{U}^{(\mathcal{G}, \vec{\alpha})}(s_{k}) \hat{\rho}_{k-1} \hat{U}^{(\mathcal{G}, \vec{\alpha}) \dagger}(s_{k}) + (1-T) \hat{\rho}_{\text{vac}}$, where the multimode gate $\hat{U}^{(\mathcal{G}, \vec{\alpha})}(s_{k})$ depends on the cluster topology, $\mathcal{G}$, the encoding values $\vec{\alpha} \equiv \left\{ \alpha_{i}, \beta_{i} \right\}_{i=1}^{N}$ and the input $s_{k}$ (see App. Sect. \ref{app-dynamics}) and $\hat{\rho}_{\text{vac}}$ is the density matrix of an $N$-mode product of vacuum states.
The cluster coupling strengths, as well as the encoding values from $\vec{\alpha}$ and the BS transmissivity are key hyperparameters for QRC to work properly in this protocol. Concretely, the coupling strengths of the resource cluster and the encoding strengths, $\vec{\alpha}$, have an energy-enhancing effect while the BS transmissivity allows tuning losses and memory \cite{GBeni2023,GBeni2024} (see App. Sect. \ref{app-fading-QRC} for details).

\section{Performance in machine learning tasks}
We now test the performance of our QRC protocol on different benchmark tasks. We have chosen a set of temporal and static tasks to display the versatility of our proposal. The XOR gate is shown both for sequential data (temporal XOR) as well as with static inputs (static XOR); then we consider the \textit{nonlinear autoregressive moving average} (NARMA) task; finally, we implement a QRC protocol for image classification on the MNIST handwritten digit dataset.

All the simulations of the quantum system have been carried out in the Heisenberg picture, using the symplectic formalism for Gaussian quantum dynamics \cite{Adesso_2014} using the Python programming language, and libraries such as NumPy \cite{numpy} and SciPy \cite{scipy} for the quantum reservoir dynamics, Scikit-learn \cite{scikit-learn} and Keras \cite{keras2015} to apply the readout optimization, and Matplotlib \cite{matplotlib} and Seaborn \cite{seaborn} for data visualization. Throughout the manuscript, we keep as fixed hyperparameters {the BS transmissivity $T = 0.4$, unless stated otherwise. Additionally, we adjust the encoding strengths $(\alpha_{i}, \beta_{i})$ from Eq. \eqref{p-prime-sk-eq} in conjunction with an increase in reservoir size, ensuring that the quantum reservoir exhibits fading memory (see App. Sect. \ref{app-fading-QRC}).}

\subsection{Temporal and static XOR tasks}
\begin{figure}[h!]
    \centering
    \includegraphics[width=\linewidth]{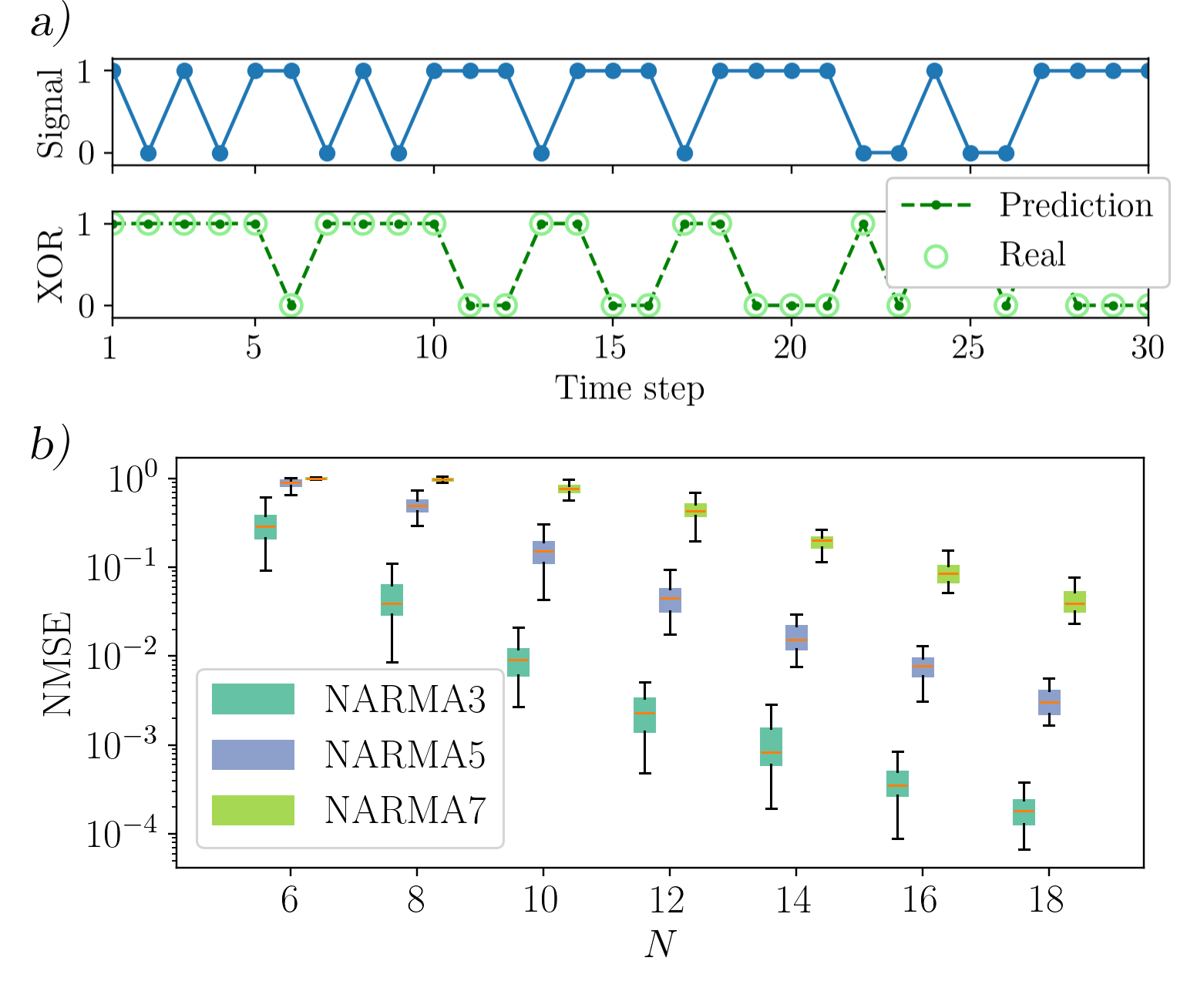}
    \caption{(a) Temporal XOR task: in the top plot (blue lines), a sample of the injected input sequence is shown as a function of the time step. In the plot below, the predicted values of the temporal XOR task are shown (dark green dashed lines) and compared to the real true values (empty circles). Training and test sets of 3000 and 500 consecutive values, respectively (and previous wash-out set of 100 values). (b) Boxplot depicting the NMSE of the NARMA task as a function of the reservoir size, $N$. Each box is calculated by taking the values from 50 independent realizations. Different colors show different NARMA delays. For a given value of the x-axis, the boxes for each delay scenario are split to avoid overlapping. Same sizes of training and test set as in (a).}
    \label{figure-2}
\end{figure}

\textit{Temporal XOR. } The first task we implement is the temporal XOR gate. We consider a random input sequence comprised of zeros and ones as can be seen in the top panel of Fig. \ref{figure-2}a. At the $k$-th time step, the target function is the XOR gate of the input $s_{k}$ and the previous one, $s_{k-1}$, so that $\bar{y}_{k} = \text{XOR}(s_{k},s_{k-1})$. It is equal to one if $s_{k} \neq s_{k-1}$ and zero otherwise. This task is relatively easy to implement in a quantum reservoir but serves as a proof of principle, as it requires non-linearity and memory. To implement this task we consider a two-mode reservoir ($N = 2$), the smallest size reservoir that can implement the task. The resource cluster is a two-mode cluster state and the measurement basis is chosen to be $\hat{p}_{1}'(s_{k}) = \hat{p}_{1} + s_{k} \hat{x}_{1}/3$ and $\hat{p}_{2}'(s_{k}) = \hat{p}_{2}$. The input is thus only encoded in one of the modes, keeping the other detector static, which adds heterogeneity to the observables and is less experimentally demanding. The initial state is in a vacuum state, $\ket{\Psi_{0}} = \ket{0} \otimes \ket{0}$ ($\hat{\rho}_{0} = \ket{\Psi_{0}}\bra{\Psi_{0}}$). As the target function is binary, a logistic regression is applied in the readout layer to the measured observables (see App. Sect. \ref{logistic-app}). The two-mode measurement based reservoir is able to implement the temporal XOR gate with a 100\% accuracy, as can be seen in Fig. \ref{figure-2}a.

\textit{Static XOR. } For the static XOR task, we take two independent inputs, $s_{1}$ and $s_{2}$, and encode them to the two-mode reservoir simultaneously. For that, we encode each input into a different detector, so that we measure the observables $\hat{p}_{i}'(s_{i}) = \hat{p}_{i} + s_{i} \hat{x}_{i}/3$ ($i = 1, 2$ being the mode label). In this case, the input inhomogeneity is not required, as we are encoding a different input to each mode. Properly speaking, this is an extreme learning machine task more than an RC one and does not require any internal memory in the reservoir \cite{opportunities}. Hence, for each input pair, the reservoir state is reset to the same initial condition, say $\ket{\Psi_{0}}$, so we remove any temporal dependence, as opposed to the temporal XOR task. The target function is in this case $\bar{y}(s_{1},s_{2}) = \text{XOR}(s_{1}, s_{2})$. As there is no sequential data and no fading memory, the initial conditions of the reservoir before the input encoding play a crucial role in the performance. Concretely, the vacuum initial condition used in the temporal XOR task only allows this protocol to get a 75\% accuracy in the static XOR. On the other hand, using a two-mode coherent state as the initial state of the form $\ket{\Psi_{0}} = \ket{\alpha_{x}} \otimes \ket{\alpha_{x}}$ (so that $\bra{\alpha_{x}} \hat{x} \ket{\alpha_{x}} = \alpha_{x}$ and $\bra{\alpha_{x}} \hat{p} \ket{\alpha_{x}} = 0$) yielded a 100\% accuracy in both the training and test set. The observed results can be attributed to the fact that, for a single teleportation step, the second-order moments of the initial vacuum state are not functions of the product $s_{1} \cdot s_{2}$ (a necessary nonlinearity for the task). However, they are functions of this product for the two-mode coherent state (see App. Sect. \ref{app-XOR} for details).

\subsection{NARMA task} 
The NARMA task requires both high memory and nonlinearity (see App. Sect. \ref{app-narma} for details on the task). It is one of the most common benchmark tasks for time series processing and has already been tested in several QRC proposals \cite{Martinez-Peña2021,Nokkala2022,GBeni2024}. Concretely, we consider the family of tasks NARMA$d$, where the parameter $d$ determines the delay of the task (See App. Sect. \ref{app-narma}). Higher delays require a higher memory from the reservoir to reproduce the task. The resource cluster states that are used in this instance have an open-chain topology,
\begin{equation} \label{chain-cluster}
    \ket{\Phi_{\text{chain}}} = \prod_{i = 1}^{N-1} \hat{C}_{Z}^{(i,i+1)} \bigotimes_{k=1}^{N} \ket{0}_{\hat{p}_{k}} ,
\end{equation}
where every mode has a link to its nearest neighbors but modes 1 and $N$ are not linked to one another. This is one of the most simple ways to get an interconnected cluster network. The reservoir is initially set to the vacuum state, as in the temporal XOR task. The engineered measurement bases are chosen to be $\hat{p}_{i}'(s_{k}) = \hat{p}_{i} + \left(\alpha_{i} s_{k} + \beta_{i} \right) \hat{x}_{i}$, where $\alpha_{i}$ and $\beta_{i}$ are uniformly distributed between $-0.2$ and $0.2$, ensuring heterogeneity to the measured observables. The reduction of the encoding values magnitude, $\alpha_{i}$ and $\beta_{i}$, compared to the XOR tasks, is to ensure the fading memory property as higher reservoir sizes are considered (see App. Sect. \ref{app-fading-QRC} for details). The performance of this task is measured using the \textit{normalized mean-square error} (NMSE) between the predicted and the target function $\text{NMSE}(\mathbf{y},\bar{\mathbf{y}}) = \left[\sum_{k=1}^{M} (y_{k}-\bar{y}_{k})^{2}\right]/\left[\sum_{k=1}^{M} \bar{y}_{k}^{2} \right]$. In Fig. \ref{figure-2}b the error in the NARMA task is shown as a function of the reservoir size, $N$, for different delays $d$. The performance improvement as the reservoir size is increased is seen for every delay. In the case of $d = 7$ the performance does not improve until $N > 10$. This is expected, as for the NARMA$d$ tasks to be faithfully reproduced, the reservoir needs to reach linear and quadratic memories up to delay $d$ \cite{Kubota2021}. It is also noteworthy that our reservoir can implement the NARMA task while remaining in a vacuum state by just measuring the second-order moments. Previous Gaussian QRC frameworks required either using amplitude encoding on coherent states \cite{Nokkala2022} or measuring fourth-order moments \cite{GBeni2024} to access the necessary nonlinearities, which is not required in this setup.

\subsection{MNIST image classification.} 
Image classification is a popular benchmark task for reservoir computers both in the classical domain \cite{Antonik2020,Nakajima2021} and in the quantum domain \cite{Sakurai2022,memristor,kornjaca2024}. We use the MNIST handwritten digit dataset consisting of 70000 hand-drawn images of 28x28 pixels representing numbers from 0 to 9 \cite{mnist_cite}. The dataset is split into 60000 training images and 10000 test images. To reduce the dimensionality of the images and make the input compatible with a numerically accessible cluster size, we use the \textit{Zoning 2} compression technique to compress every image in the dataset to 14x14 pixels (see App. Sect. \ref{app-MNIST}) as in other optical reservoir computing frameworks \cite{Antonik2020}.

\begin{figure}[h!]
    \centering
    \includegraphics[width=\linewidth]{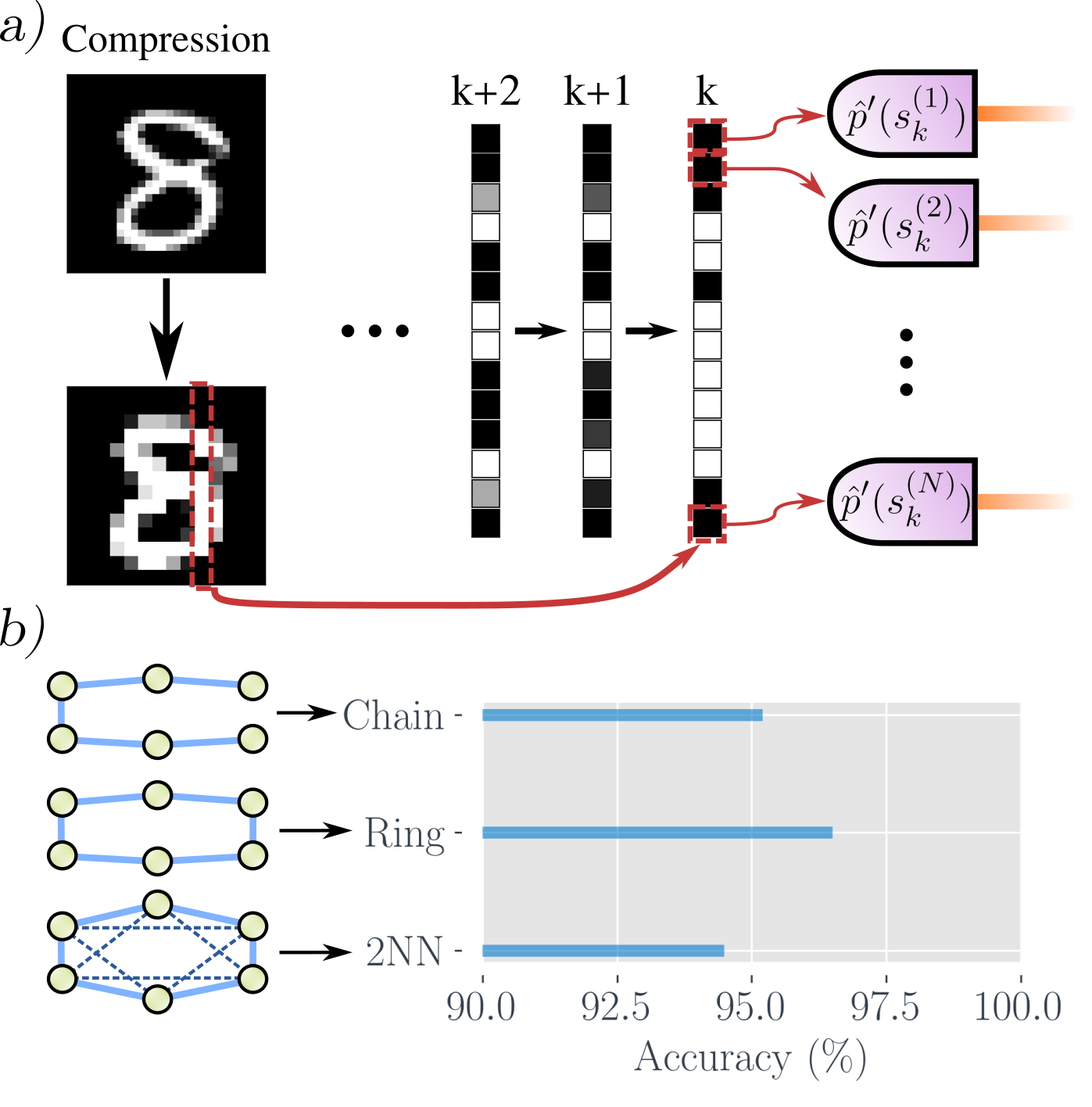}
    \caption{(a) Scheme of the encoding protocol for the MNIST task. (b) The horizontal histogram plot shows the test accuracy (x-axis) for different resource cluster topologies (y-axis). The accuracy was determined as the highest value among $50$ epochs. The six-mode clusters shown on the left are visual representations of the used topologies, the clusters employed in the simulations contained 14 modes. The dashed lines of the 2NN couplings depict their smaller coupling strength, as explained in the main text.}
    \label{figure-3}
\end{figure}

While image classification is a static task, it can be executed by exploiting the device's internal memory. The advantage is that relatively small reservoirs, with respect to image size, can achieve the needed performance. For this purpose, each image is sliced as pixels arranged in an ordered sequence of columns, and at each time step the column vector is injected as input in the reservoir computing as illustrated in Fig.\ref{figure-3}a. This substantially diminishes the minimum reservoir size required for processing the input, which can be introduced by dividing the compressed image into columns (14-dimensional vectors). This technique has already been used for QRC protocols \cite{memristor}.
The described encoding protocol for the MNIST task with the image embedded in the measured quadratures is shown in Fig. \ref{figure-3}a.
Concretely, every single pixel from each column is encoded into an independent mode, thus requiring a reservoir with a size equal to the number of pixels in each column, $N = 14$. By labeling every input pixel as $s_{k}^{(i)}$, where $k$ is the (temporal) column label and $i$ is the row (mode) label ($k, i = 1, 2, \dots, 14$), we set the teleportation measurement basis to $\hat{p}_{i}'(s_{k}^{(i)}) = \hat{p}_{i} + 0.1 \left( s_{k}^{(i)} + 1\right) \hat{x}_{i}$. As in the static XOR task, a homogeneous encoding is sufficient.

We consider three different cluster topologies for this specific task, as they yield different performance results (unlike in the NARMA task), namely, a chain, as in Eq. \eqref{chain-cluster}; a ring topology, $\ket{\Phi_{\text{ring}}} = \hat{C}_{Z}^{(1,N)} \ket{\Phi_{\text{chain}}}$, in which a link is added connecting the first and last modes; we also consider a second nearest neighbors (2NN) topology, $\ket{\Phi_{\text{2NN}}} = \hat{C}_{Z}^{(1,N-1)}(\xi) \hat{C}_{Z}^{(2,N)}(\xi)\prod_{i=1}^{N-2} \hat{C}_{Z}^{(i,i+2)}(\xi) \ket{\Phi_{\text{ring}}}$, where we add 2NN links (with strength equal to $\xi = 0.1$) to the ring cluster. The smaller strength of the 2NN couplings improves the fading memory of the reservoir, as the number of mode couplings is substantially increased. The BS transmissivities are reduced to $T = 0.3$ for this task for the same reason (for all network topologies). The readout observables of the quantum reservoir are also sequentially taken at every time step, as in any temporal task. This means $105$ observables (the number of second-order moments)  are measured at each time step for a total of $14$ time steps, which yields a total of $1470$ observables. During post-processing, the observables from the first time step were removed, as the first column of every image in the dataset was blank. A softmax regression was implemented in the readout layer for the classification task (see App. Sect. \ref{softmax-app}). The accuracy obtained on the test set is shown in Fig. \ref{figure-3}b for the three cluster topologies. The ring topology got the best performance, with 96.5\% accuracy, while the 2NN topology got the worst accuracy (below 95\%). Here, cluster network topology played an important role, and we could expect better performance for task-optimized topologies.

\section{Discussion}
In the current era of data and communications, photonic quantum technologies have the potential to significantly enhance the speed and efficiency of information processing. When considering quantum resources, CV cluster states are gaining prominence due to experimental advances in their on-demand generation and their versatility for application across multiple domains. We have shown that our quantum photonic platform is able to process classical information using CV cluster states as resources for machine learning purposes. Concretely, we introduce a design for measurement based quantum reservoir computing that significantly extends the potential of photonics in data processing. Here cluster states and teleportation are essential resources that enable reservoir operations in an entangled state while input encoding and monitoring are realized locally. 
The addition of BS after the teleportation measurements allows the quantum reservoir to be continuously monitored to extract information from the quantum reservoir without the need to restart the experiment and thus preserving the memory of the previous inputs \cite{mujal_meas,GBeni2023}. Interestingly, beyond the needed monitoring, this is also an essential mechanism to enable the reservoir computing operation with a fading memory \cite{GRIGORYEVA2018495}.
We have tested this platform in several temporal and static tasks, reaching good performances in every one of them with minimal setup adjustments and addressing the presence of internal memory, opening the way to sequential (in-memory) processing with this setting. The effects of size scaling and cluster topology were briefly explored in some tasks. This design goes beyond classical photonics for machine learning \cite{PhotRC_book,Shastri} by incorporating entanglement and teleportation, as well as MBQC in qubits settings \cite{MB_VQE} being based on CV, and targeting time series processing, Additionally, it features internal memory, expanding the capabilities of extreme machine learning \cite{suprano}.

Our work opens up several avenues for research of measurement based machine learning and reservoir computing. The capacity to drive the reservoir into a quantum network state by local operations makes the protocol suitable for quantum distributed computing \cite{Koudia_2024}, where the computation is performed among different separated parties \cite{caleffi2022}. The proposal can be adapted to different experimental constraints addressing for instance the effect of statistical noise, entanglement and finite squeezing in the protocol. Furthermore, this design can also be adapted to variational settings where the choice of the measurements also serves to optimize the reservoir for a given task \cite{MB_VQE}. Going beyond Gaussian states is necessary for a large range of quantum information processing tasks  \cite{Bartlett02,Mari12,Rahimi-Keshari16}.  Our platform may thus also benefit from the use of non-Gaussian operations in the teleportation protocol, enhancing the available Hilbert space to a greater extent and increasing expressivity.

In conclusion, our findings highlight the potential of CV cluster states as a versatile resource for measurement based QRC and more in general for machine learning and broader quantum computing applications. This work establishes a foundation for experimental implementations and future research into optimizing cluster state topology, scaling system size, and integrating advanced quantum operations to enhance performance and explore novel computational paradigms.

\begin{acknowledgments}
We acknowledge the Spanish State Research Agency, through the Mar\'ia de Maeztu project CEX2021-001164-M,  {the COQUSY projects PID2022-140506NB-C21 and -C22, and the INFOLANET project PID2022-139409NB-I00, all funded by
MICIU/AEI/10.13039/501100011033, and by ERDF, EU}; MINECO through the QUANTUM SPAIN project, and EU through the RTRP - NextGenerationEU within the framework of the Digital Spain 2025 Agenda.  GLG is funded by the Spanish  Ministerio de Educaci\'on y Formaci\'on Profesional/Ministerio de Universidades and co-funded by the University of the Balearic Islands through the Beatriz Galindo program (BG20/00085). JGB is funded by the Conselleria d’Educació, Universitat i Recerca of the Government of the Balearic Islands with grant code FPI/036/2020. VP acknowledges the   European   Research Council under the Consolidator Grant COQCOoN (Grant No.  820079).

\textbf{Conflicts of interest:} The authors declare no conflicts of interest.

\textbf{Data availability:} All data needed to evaluate the conclusions in the paper are present in the paper and/or the Appendix. A code sample can be found at \cite{github}.
\end{acknowledgments}

\appendix
\section{Recall of  single mode and multimode CV quantum teleportation} \label{app-QTP}
In this section, we explain in greater detail the teleportation procedure used in the main text. First, we consider the single-mode teleportation circuit; then, we move to the two-mode setting where two-mode gates are introduced; finally, we expand this protocol to larger multimode settings.
\subsection{Single mode CV teleportation}
\begin{figure}
    \centering
    \includegraphics[width=\linewidth]{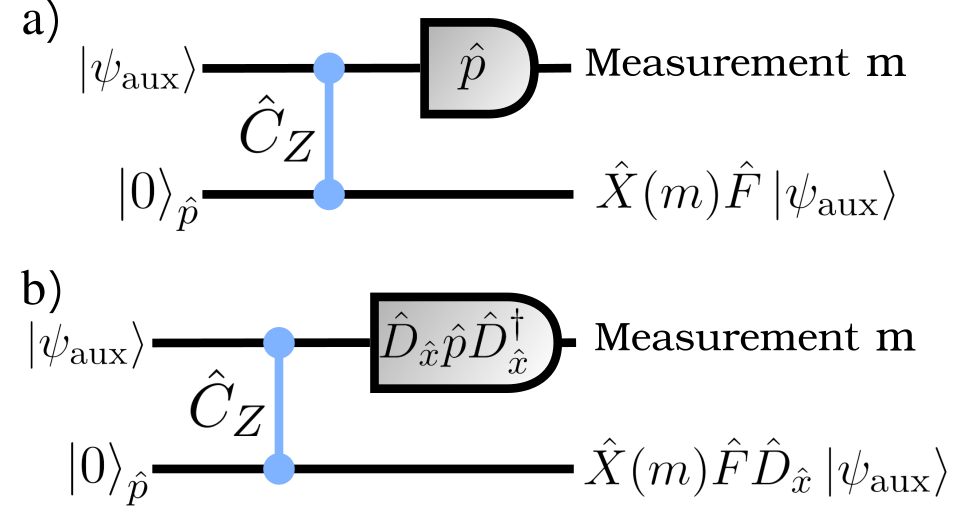}
    \caption{Scheme for a single mode CV teleportation step where: (a) there is no basis change and a simple teleportation is applied; (b) the measurement basis is tuned and thus a quantum gate is applied to the resulting state.}
    \label{app-fig1}
\end{figure}

In this section, we continue the single-mode teleportation circuit discussed in the main text. For an auxiliary state, $\ket{\psi_{\text{aux}}}$, which we want to teleport, we use a $p$-quadrature eigenstate, $\ket{0}_{\hat{p}}$, as a resource and couple both using an ideal $\hat{C}_{Z}$ gate, $\hat{C}_{Z}^{(A,B)} = \exp\left[ -i \hat{x}_{A} \otimes \hat{x}_{B} \right]$. Then, a detector measuring the $\hat{p}$ quadrature of the auxiliary state is placed, performing the teleportation \cite{Gu2009, Ukai2015_ch4} (as depicted in Fig. \ref{app-fig1}a). The teleported state, $\ket{\psi_{\text{QTP}}}$, becomes
\begin{equation}
    \ket{\psi_{\text{QTP}}} = \hat{X}(m) \hat{F} \ket{\psi_{\text{aux}}} \ ,
\end{equation}
where $\hat{F} = \exp\left[ i \pi \left( \hat{x}^{2} + \hat{p}^{2} \right)/2 \right]$ and $\hat{X}(m) = \exp \left( -i m \hat{p} \right)$ ($m$ being the measurement result). So we end up with a conditional state that is proportional to the initial auxiliary states plus extra gates that do not depend on the initial state, as typical in quantum teleportation protocols.

Let us consider the effect of applying a gate $\hat{D}_{\hat{x}} = e^{i f(\hat{x})}$. As this family of gates commutes with the $\hat{C}_{Z}$ gate \cite{Gu2009}, applying it to the initial state $\ket{\psi_{0}}$ before the teleportation protocol is equivalent to applying it after the $\hat{C}_{Z}$ coupling and before the teleportation measurement \cite{Ukai2015_ch4}.
Furthermore, applying the $\hat{D}_{\hat{x}}$ gate before the measurement is equivalent to performing detection in the transformed basis $\hat{p}' = \hat{D}_{\hat{x}} \hat{p} \hat{D}_{\hat{x}}^{\dagger}$. In Fig. \ref{app-fig1}b, a scheme of the circuit with the engineered measurement, $\hat{p}'$, is shown. In this case, the resulting state can be written as
\begin{equation}
    \ket{\psi_{\text{QTP}}} = \hat{X}(m) \hat{F} \hat{D}_{\hat{x}} \ket{\psi_{\text{aux}}} \ .
\end{equation}
We have thus proven that, due to the mathematical properties of the $\hat{C}_{Z}$ gate, the family of gates $\hat{D}_{\hat{x}} = e^{i f(\hat{x})}$ can be deterministically applied to any auxiliary state by performing the teleportation circuit and measuring the observable $\hat{D}_{\hat{x}} \hat{p} \hat{D}_{\hat{x}}^{\dagger}$. In the main text the quadratic gate, $\hat{D}_{\hat{x}} = e^{-i s \hat{x}^{2}}$, is considered, which corresponds to measuring the observable $\hat{p} + s \hat{x}$. This measurement is experimentally accessible by phase rotation and field amplification \cite{Miwa2009}.

The stochastic gate, $\hat{X}(m)$, can be removed by applying its inverse, $\hat{X}(-m)$, to $\ket{\psi_{\text{QTP}}}$. For a multiple-step teleportation, each stochastic displacement $\hat{X}(m)$ has to be either removed sequentially with conditioned local displacements between teleportation sequences or in postprocessing by keeping track and counting every sequence's contribution \cite{Gu2009}. In the main text, we assume these conditioned corrections will be made either way during the QRC protocol.
\subsection{Two-mode CV teleportation}
\begin{figure}
    \centering
    \includegraphics[width=0.65\linewidth]{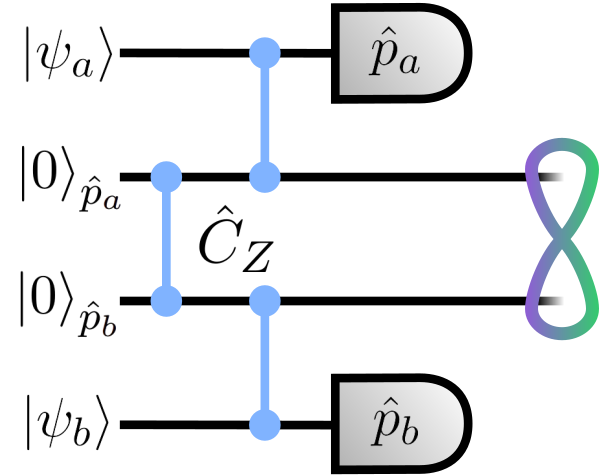}
    \caption{Scheme for a two-mode CV teleportation step.}
    \label{App-fig2}
\end{figure}

For a two-mode teleportation circuit in which two-mode gates are implemented, we consider an initial auxiliary product state, $\ket{\Psi_{\text{aux}}} = \ket{\psi_{a}} \otimes \ket{\psi_{b}}$. The resource state is now a two-mode cluster state, $\ket{\Phi_{\text{cl}}} = \hat{C}_{Z}^{(a,b)}\ket{0}_{\hat{p}_{a}} \otimes \ket{0}_{\hat{p}_{b}}$ (here we consider an ideal $\hat{C}_{Z}$ gate, strength $\xi = 1$, coupling both $p$-quadrature eigenstates, but the following explanation can be done with arbitrary coupling strengths). Applying a local circuit to both auxiliary modes implements the two-mode circuit (see Fig. \ref{App-fig2}). It is equivalent to having two single-mode circuits, one for $\ket{\psi_{a}}$ and the other for $\ket{\psi_{b}}$, but using two $\ket{0_{\ket{p}}}$ states that are entangled with a $\hat{C}_{Z}$ coupling, becoming a two-mode cluster state. 
The main intuition of why this two-mode circuit implements an entangling gate between the auxiliary states is as follows. All $\hat{C}_{Z}$ gates in the circuit commute with each other, so that applying the $\hat{C}_{Z}$ gate between the $p$-quadrature eigenstates before the teleportation circuits is mathematically identical to applying it at the end of the protocol (after the teleportation) \cite{Ukai2015_ch5}. Because of this, it can be easily proven that the teleported state is
\begin{equation} \label{qtp-2mode-disp}
\begin{aligned}
    \ket{\Psi_{\text{QTP}}} &= \left[\hat{X}(m_{a}) \hat{Z}(m_{b}) \otimes \hat{X}(m_{b}) \hat{Z}(m_{a}) \right] \hat{C}_{Z}^{(a,b)} \\ 
    &\times \hat{F}^{(2)}\ket{\Psi_{\text{aux}}} \ ,
\end{aligned}
\end{equation}
where $\hat{F}^{(2)} = \hat{F} \otimes \hat{F}$, $\hat{Z}(m) = e^{i m \hat{x}}$ and the magnitude $m_{a}$ ($m_{b}$) is the result of the measurement of $\hat{p}_{a}$ ($\hat{p}_{b}$). After removing the stochastic displacements, the state becomes
\begin{equation}
    \ket{\Psi_{\text{QTP}}} = \hat{C}_{Z}^{(a,b)} \hat{F}^{(2)} \ket{\Psi_{\text{aux}}} \ ,
\end{equation}
where a $\hat{C}_{Z}$ gate between both auxiliary modes have been applied. This protocol can implement engineered single-mode gates by measuring the observables $\hat{p}'_{a} = \hat{D}_{\hat{x}}^{(a)} \hat{p}_{a} \hat{D}_{\hat{x}}^{(a) \dagger}$ and $\hat{p}'_{b} = \hat{D}_{\hat{x}}^{(b)} \hat{p}_{b} \hat{D}_{\hat{x}}^{(b) \dagger}$, where $\hat{D}_{x}^{(a)} = e^{i f_{a}(\hat{x})}$ and $\hat{D}_{x}^{(b)} = e^{i f_{b}(\hat{x})}$ (they can be mode-dependent operators, i.e. $f_{a}(\hat{x}) \neq f_{b}(\hat{x})$). In that case, the teleported state is $\ket{\Psi_{\text{QTP}}} \propto \left(\hat{D}_{x}^{(a)} \otimes \hat{D}_{x}^{(b)}\right) \ket{\Psi_{\text{aux}}}$, including the same operators as in Eq. \eqref{qtp-2mode-disp}.

\subsection{Multimode CV teleportation}
The ideas from the previous section can be extended to an arbitrarily large multimode teleportation protocol. We consider a separable $N$-mode auxiliary state $\ket{\Psi_{\text{aux}}} = \bigotimes_{i=1}^{N} \ket{\psi_{i}}$. For this multimode setting we require an $N$-mode cluster state defined as in Eq. \ref{cluster-eq} of the main text. For the $N > 2$ case, the graph topology $\mathcal{G}$ of the cluster state is a factor to consider. We perform local teleportation circuits as in Fig. \ref{app-fig1} where every auxiliary mode gets coupled to every cluster mode one-to-one.  Analogously to Eq. \eqref{qtp-2mode-disp}, if we consider engineered measurements, $\hat{p}_{i}' = e^{i f_{i}(\hat{x}_{i})} \hat{p}_{i} e^{-i f_{i}^{*}(\hat{x}_{i})}$, and remove the undesired stochastic displacements we end up with the expression
\begin{equation} \label{psi-QTP-N-eq}
    \ket{\Psi_{\text{QTP}}} = \hat{C}_{Z}^{(\mathcal{G})} \hat{F}^{(N)} \hat{D}_{\hat{x}}^{(N)}\ket{\Psi_{\text{aux}}} \ ,
\end{equation}
where $\hat{C}_{Z}^{(\mathcal{G})}$ comprises the $\hat{C}_{Z}$ operators present in the cluster (as defined in Eq. \eqref{cluster-eq}), $\hat{F}^{(N)} = \bigotimes_{i=1}^{N} \hat{F}$ and $\hat{D}_{\hat{x}}^{(N)} = \bigotimes_{i=1}^{N} e^{i f_{i}(\hat{x}_{i})}$.

In the following sections, we move to the Heisenberg picture, describing the dynamics of the quadratures. In this picture, all Gaussian operators become symplectic matrices, which simplifies the expressions \cite{Adesso_2014}. If we consider a quadrature vector from an $N$-mode state
\begin{equation}
    \hat{\mathbf{Q}} = \left( 
    \begin{array}{c}
         \hat{x}_{1} \\
         \hat{p}_{1} \\
         \vdots \\
         \hat{x}_{N} \\
         \hat{p}_{N}
    \end{array}
    \right) \ ,
\end{equation}
then we can rewrite Eq. \eqref{psi-QTP-N-eq} as
\begin{equation} \label{Q-QTP-N-eq}
    \hat{\mathbf{Q}}_{\text{QTP}} = C_{Z}^{(\mathcal{G})} F^{(N)} D^{(N)}_{\hat{x}} \hat{\mathbf{Q}}_{\text{aux}} \ .
\end{equation}
In this case, the operators have been swapped for their symplectic representation. Now $C_{Z}^{(\mathcal{G})}$ is built from a $2N \times 2N$ diagonal matrix with the added terms
\begin{equation}
    \left[C_{Z}^{(\mathcal{G})}\right]_{2i+1,2j} = \left\{
    \begin{aligned}
        \xi_{ij} \quad & \text{if} \ (i,j) \in \mathcal{G} \\
        0 \quad & \text{otherwise}
    \end{aligned}
    \right. \ ,
\end{equation}
where we take $\left[C_{Z}^{(\mathcal{G})}\right]_{2j+1,2i} = \left[C_{Z}^{(\mathcal{G})}\right]_{2i+1,2j}$, $\xi_{ij}$ being the coupling strength between modes $i$ and $j$. Moreover, $F^{(N)} = \bigoplus_{i=1}^{N} \left( 
\begin{array}{cc}
    0 & -1 \\
    1 & 0
\end{array}
\right)$. If we consider engineered quadratic gates, $f_{i}(\hat{x}_{i}) = -s_{i} \hat{x}_{i}^{2}$, as in the main text, then $D^{(N)}_{\hat{x}} = \bigoplus_{i=1}^{N} \left( 
\begin{array}{cc}
    1 & 0 \\
    s_{i} & 1
\end{array}
\right)$, where $s_{i}$ is mode-dependent ($i = 1, 2, \dots, N$).

\section{Reservoir Computing} \label{app-RC}
In this section, we will explain the RC scheme in more detail. Every RC scheme can be separated into three steps/layers: the input layer, the reservoir dynamics, and the readout layer. At time step $k$ a vector $\mathbf{s}_{k}$ from the time series signal, $\{\mathbf{s}_{1},\mathbf{s}_{2},\dots,\mathbf{s}_{L} \}$, is encoded and fed into the dynamical system, also called \textit{reservoir}. Each $\mathbf{s}_{k}$ can be a vector, as in the case of the MNIST task, or a scalar, as in the case of the NARMA task. The length of the sequence, $L$, denotes the number of training samples, or training steps, to perform the learning protocol. After the input encoding, the $D_{R}$ reservoir degrees of freedom, $\mathbf{v}_{k} \in \mathcal{R}^{D_{R}}$, evolve through a nonlinear mapping and are measured as outputs. The reservoir state, $\mathbf{v}_{k}$, is recorded in the readout layer. After this step, the protocol continues for the following input, $\mathbf{s}_{k+1}$, at time step $k+1$. Once all the input sequence has been encoded, a regression is applied to the recorded observables in the readout layer to yield the output functions $\{\mathbf{y}_{1}, \mathbf{y}_{2}, \dots, \mathbf{y}_{L}\}$. Every function $\mathbf{y}_{k}$ is built from the reservoir observables, $\mathbf{v}_{k}$, and a set of tunable weights. These weights are optimized during the training stage for the output to converge to a desired target function $\bar{\mathbf{y}}_{k}$.

We can write the dynamical mapping of the reservoir degrees of freedom as $\mathbf{v}_{k} = \mathcal{H} \left( \mathbf{v}_{k-1}, \mathbf{s}_{k} \right)$ ($k \in \mathbb{Z}$), in which we need $\mathcal{H}$ to be a nonlinear mapping of the input $\mathbf{s}_{k}$. This mapping will remain fixed throughout the whole protocol.

\subsection{Linear regression.}
The most standard RC practice is to use a simple linear regression at the output layer, which implies building the output function as a linear function of the reservoir observables. It can be written as
\begin{equation}
\begin{aligned}
    y_{k} &= w_{0} + \mathbf{w}^{\top} \mathbf{v}_{k} \\
    &= \left( 1 , \mathbf{v}_{k}^{\top} \right) \left( \begin{array}{c}
       w_{0}  \\
       \mathbf{w}
    \end{array} \right) \ ,
\end{aligned}
\end{equation}
with the $(D_{R} + 1)$-dimensional vector $\mathbf{W} = \left(w_{0}, \mathbf{w}^{\top}\right)^{\top}$ containing the training weights (we add a constant bias term $w_{0}$). For a given set of $L$ inputs that we feed to the reservoir, we can define the following matrices
\begin{equation}
    V = \left(
    \begin{array}{cc}
        1 & \mathbf{v}_{1}^{\top} \\
        1 & \mathbf{v}_{2}^{\top} \\
        \vdots & \vdots \\
        1 & \mathbf{v}_{L}^{\top}
    \end{array}
    \right) \quad ; \quad  \mathbf{y} = \left( \begin{array}{c}
        y_{1}  \\
        y_{2} \\
        \vdots \\
        y_{L}
    \end{array} \right) \ ,
\end{equation}
so that the following equation holds $\mathbf{y} = V \mathbf{W}$. After choosing a given target function that we want our reservoir to reproduce, $\bar{\mathbf{y}}$, we want to find the weights that minimize the \textit{mean-square error} (MSE) of the predicted $\mathbf{y}$ and the target. The loss function to minimize in this case is $\mathcal{L}_{\text{MSE}} = \frac{1}{L}\sum_{k=1}^{L} (y_{k} - \bar{y}_{k})^{2}$. The optimal set of weights for this condition to hold are the ones obtained through the following procedure, \cite{LUKOSEVICIUS2009127}:    $\mathbf{W}_{\text{opt}} = V^{\text{MP}} \bar{\mathbf{y}}$, where $V^{MP} = \left(V^{\top} V\right)^{-1} V^{\top}$ is the Moore-Penrose inverse of $V$. The higher the value of $L$ (size of the training sequence) the more precise our estimation of the optimal weights will usually be.

\subsection{Logistic regression}\label{logistic-app}
The standard ML binary classification practice is applying a logistic regression \cite{cox1958regression}. A logistic regression implies a logistic function applied to the readout layer, so at time step $k$ we define the readout function $\mathcal{P}_{k} = \left[1 + \exp\left( -w_{0} - \mathbf{w}^{\top} \mathbf{v}_{k} \right)\right]^{-1}$.
The loss function that we want to minimize in this case is the \textit{binary cross-entropy} (BCE), which can be written as
\begin{equation}
    \mathcal{L}_{\text{BCE}} = - \frac{1}{L} \sum_{k=1}^{L}\left[ \bar{y}_{k} \log(\mathcal{P}_{k}) + (1-\bar{y}_{k}) \log(1-\mathcal{P}_{k}) \right] \ .
\end{equation}
The minimization of this loss function is not as straightforward as in the linear regression case, and it is usually achieved via numerical methods such as gradient descent \cite{gradient_methods_book}. Once the weights have been optimized during the training set, the predicted functions $y_{k} = 1$ iff $\mathcal{P}_{k} > 1/2$ and zero otherwise.

\subsection{Softmax regression} \label{softmax-app}
The softmax regression is an extension of the logistic regression suitable for more than two classes. Given a set of $M$ classes, that we label from 1 to $M$, we compute the set of $M$ readout functions $\sigma_{k}^{(p)} = \exp(\mathcal{Z}_{k}^{(p)})/\sum_{i=1}^{M} \exp(\mathcal{Z}_{k}^{(i)})$,
where $\mathcal{Z}_{p} = w_{0}^{(p)} + \mathbf{w}_{(p)}^{\top} \mathbf{v}_{k}$, $p$ denoting the class label (in this case, there is a set of weights for each class). Each function $\sigma_{k}^{(p)}$ returned at the readout layer can be interpreted as the probability that the input image belongs to label $p$. We optimize the weights by minimizing the following loss function 
\begin{equation}
    \mathcal{L} = -\frac{1}{L}\sum_{k=1}^{L} \sum_{p=1}^{M} \bar{y}_{k}^{(p)} \log \sigma_{k}^{(p)} \ ,
\end{equation}
where we consider that, for the $k$-th term in the dataset, $\bar{y}_{k}^{(p')} = 1$ if it belongs to class $p'$ and zero otherwise. This optimization is also done with numerical methods, as in the case of the logistic regression. The predicted class for each different term in the dataset is the one with the highest $\sigma_{k}^{(p)}$ value.

\section{Quantum Reservoir Computing dynamics} \label{app-dynamics}
The reservoir dynamics can be summarized as follows:
\begin{itemize}
    \item The reservoir is coupled to the resource cluster state via $\hat{C}_{Z}$ gates.
    \item The reservoir modes are measured on an engineered basis and applying a teleportation protocol, so the resulting state has an engineered multimode gate applied to it.
    \item The resulting state goes through a BS coupled to the vacuum, resulting in two output paths.
    \item One of the paths gets measured via homodyne detection (this is true for every mode).
\end{itemize}
Applying Eq. \eqref{Q-QTP-N-eq}, we can obtain the covariance matrix and the first-order moment vector \cite{Adesso_2014} after the engineered teleportation:
\begin{align}
    \Gamma_{\text{QTP}}^{(k)} &=U^{(\mathcal{G}, \vec{\alpha})}_{\mathbf{s_{k}}} \Gamma^{(k)} U^{(\mathcal{G}, \vec{\alpha}) \top}_{\mathbf{s_{k}}} \\
    \mathbf{R}_{\text{QTP}}^{(k)} &= U^{(\mathcal{G}, \vec{\alpha})}_{\mathbf{s_{k}}} \mathbf{R}^{(k)} \ ,
\end{align}
where $U^{(\mathcal{G}, \vec{\alpha})}_{\mathbf{s_{k}}} = C_{Z}^{(\mathcal{G})} F^{(N)} D_{\mathbf{s_{k}}}^{(\vec{\alpha})}$ is the resulting engineered multimode gate. In this case, 
\begin{equation}
    D_{\mathbf{s_{k}}}^{(\vec{\alpha})} = \bigoplus_{i=1}^{N} \left(
\begin{array}{cc}
    1 & 0 \\
    \alpha_{i} s_{k}^{(i)} + \beta_{i} & 1
\end{array}
\right)
\end{equation}
is the symplectic matrix corresponding to the applied local quadratic gates. The mode-dependent parameter pairs $(\alpha_{i},\beta_{i})$ ($i = 1, \dots, N$) constitute the encoding vector, $\vec{\alpha}$. The $C_{Z}^{(\mathcal{G})}$ and $F^{(N)}$ gates are equivalent to those found in Eq. \eqref{Q-QTP-N-eq}. The BS symplectic matrix can be written as
\begin{equation}
    B(T) = \left(
    \begin{array}{cc}
        \sqrt{T} I_{2N} & \sqrt{1-T} I_{2N} \\
        -\sqrt{1-T} I_{2N} & \sqrt{T} I_{2N}
    \end{array}
    \right) \ ,
\end{equation}
where $I_{2N}$ is the $2N \times 2N$ identity matrix. The resulting state after the BS coupling with the vacuum can thus be written as
\begin{align}
    \Gamma_{\text{BS}}^{(k)} &= B(T) \left[\Gamma_{\text{QTP}}^{(k)} \oplus I_{2N}\right] B(T)^{\top} \\
    \mathbf{R}_{\text{BS}}^{(k)} &= B(T) \left( 
    \begin{array}{c}
        \mathbf{R}_{\text{QTP}}^{(k)}   \\
        \mathbf{0}_{2N} 
    \end{array} \right) \ ,
\end{align}
where $\mathbf{0}_{2N}$ is a $2N$-dimensional vector containing zeros. Measuring one of the paths will induce a back-action effect on the remaining state. However, performing an ensemble average of any Gaussian conditional state with no added conditioned feedback or post-selection erases the effect of the quantum back-action \cite{GBeni2023}. Thus, the remaining quantum state can be modeled as a mixed state of the teleported state and the vacuum, with covariance matrix and first-order moments equal to
\begin{align} \label{CM_R}
    \Gamma_{R}^{(k+1)} &= T U^{(\mathcal{G}, \vec{\alpha})}_{\mathbf{s_{k}}} \Gamma_{R}^{(k)} U^{(\mathcal{G}, \vec{\alpha}) \top}_{\mathbf{s_{k}}}
    + (1-T) I_{2N} \\ \label{firstmoments_R}
    \mathbf{R}^{(k+1)} &= \sqrt{T} U^{(\mathcal{G}, \vec{\alpha})}_{\mathbf{s_{k}}} \mathbf{R}^{(k)} \ ,
\end{align}
where the $k+1$ superscript is adopted because this state will be fed to the protocol at the next time step. On the other hand, if we consider a homodyne detection of the $x$-quadratures of every mode, the measured state will yield
\begin{align} \label{CM_meas_eq}
    \Gamma_{\text{meas}}^{(k)} &= \text{Tr}_{\hat{\mathbf{p}}} \left[ (1-T) U^{(\mathcal{G}, \vec{\alpha})}_{\mathbf{s_{k}}} \Gamma_{R}^{(k)} U^{(\mathcal{G}, \vec{\alpha}) \top}_{\mathbf{s_{k}}}
    + T I_{2N} \right]  \\ \label{R_meas_eq}
    \mathbf{R}_{\text{meas}}^{(k)} &= -\sqrt{1-T} \text{Tr}_{\hat{\mathbf{p}}} \left[ U^{(\mathcal{G}, \vec{\alpha})}_{\mathbf{s_{k}}} \mathbf{R}^{(k)} \right] \ .
\end{align}
The notation $\text{Tr}_{\hat{\mathbf{p}}}[ \cdot ]$ is not to be understood as the usual partial trace of the symplectic matrix inside, but as tracing out $p$-quadrature degrees of freedom. For an arbitrary $2N$-dimensional covariance matrix, $\Gamma$, tracing out these components would yield the $N$-dimensional matrix with components $\text{Tr}_{\hat{\mathbf{p}}}[\Gamma]_{ij} = \langle \hat{x}_{i} \hat{x}_{j} \rangle - \langle \hat{x}_{i} \rangle \langle \hat{x}_{j} \rangle$. From Eqs. \eqref{CM_meas_eq} and \eqref{R_meas_eq} we can easily compute the measured second-order moment matrix
\begin{equation} \label{2nd-moments-eq}
    O_{\text{2nd}}^{(k)} = \Gamma_{\text{meas}}^{(k)} + \mathbf{R}_{\text{meas}}^{(k)} \mathbf{R}_{\text{meas}}^{(k) \top} \ ,
\end{equation}
whose components are taken as the readout observables for the QRC protocol.

\subsection{Fading memory property of the QRC mapping} \label{app-fading-QRC}
In this subsection, we discuss briefly the procedure we have used to ensure the fading memory condition and echo state property \cite{GRIGORYEVA2018495} of our quantum reservoir. The mathematical conditions that quantum reservoirs undergoing linear dynamics (in the Heisenberg picture) have to fulfill are well known \cite{Nokkala2021}, as is the case of Gaussian quantum reservoirs \cite{Nokkala2021,Nokkala2022,GBeni2023,GBeni2024}. An RC protocol implemented in a generic linear dynamical system can be written as $\mathbf{v}_{k+1} = A \mathbf{v}_{k} + \mathbf{f}(\mathbf{s_{k}})$, where $\mathbf{v}_{k}$ stands for the reservoir state vector at time step $k$, $\mathbf{f}(\cdot)$ is a nonlinear vector function of the input $\mathbf{s_{k}}$ and $A$ is a $D_{R} \times D_{R}$ matrix ($D_{R}$ being the reservoir state dimensionality). Then, the fading memory of the reservoir is ensured iff $\rho(A) < 1$, where $\rho(\cdot)$ stands for the spectral radius of a matrix \cite{Nokkala2021}. This condition ensures that the reservoir implements a contractive map that forgets its initial condition. From Eqs. \eqref{CM_R} and \eqref{firstmoments_R} we can infer that the matrix we have to tackle is
\begin{equation} \label{A_raw}
\begin{aligned}
A\left(\mathcal{G},\vec{\alpha},T,\mathbf{s_{k}} \right) &= \sqrt{T} U_{\mathbf{s_{k}}}^{(\mathcal{G},\vec{\alpha})} \\
&= \sqrt{T} C_{Z}^{\mathcal{G}} F^{(N)} D_{\mathbf{s_{k}}}^{(\vec{\alpha})} \ .
\end{aligned}
\end{equation}
In our case, the dynamical matrix $A$ is a function of the cluster topology, $\mathcal{G}$, the encoding parameters, $\vec{\alpha} \equiv \left\{\alpha_{i}, \beta_{i}\right\}_{i=1}^{N}$ and the BS transmissivity, $T$. It is also dependent on the input vector, $\mathbf{s_{k}}$, which means that the spectral radius will be dependent on the time step, which contrasts with previous Gaussian QRC implementations \cite{Nokkala2021,GBeni2023}. Furthermore, the symplectic matrix $D_{\mathbf{s_{k}}}^{(\vec{\alpha})}$ from Eq. \eqref{A_raw} is not unitary, which means it does not conserve the energy of the quantum state. Indeed, the amount of energy provided by the transformation depends on the absolute value of the input (as the gate is quadratic). To standardize the fading memory procedure we need to get rid of the time step dependency. We do that by swapping the input, $\mathbf{s_{k}}$, with the maximum possible one, max$|\mathbf{s_{k}}|$. The vector max$|\mathbf{s_{k}}|$ contains the maximum absolute values that an input can have from a certain input dataset. For example, if the inputs are taken from a uniform distribution between -1 and 1, then max$|\mathbf{s_{k}}|$ would be a vector containing ones. With this, we are able to write the time-step-independent matrix
\begin{equation}
     A_{\text{max}}\left(\mathcal{G},\vec{\alpha}, T \right) \equiv \sqrt{T} C_{Z}^{\mathcal{G}} F^{(N)} D_{\text{max}\left|\mathbf{s_{k}}\right| }^{(\vec{\alpha})} \ .
\end{equation}
If we define its spectral radius as
\begin{equation}
    \lambda(\left(\mathcal{G},\vec{\alpha}, T \right) = \rho\left[ A_{\text{max}}\left(\mathcal{G},\vec{\alpha}, T \right) \right] \ ,
\end{equation}
then the condition $\lambda\left(\mathcal{G},\vec{\alpha}, T \right) < 1$ ensures both the fading memory condition and the echo state property. It is important to note that, even though there is no input dependency, the other hyperparameters have to be balanced to ensure this condition. Concretely, energy increases come from the cluster topology (more $\hat{C}_{Z}$ couplings imply more energy) and the encoding vector, while the BS determines the energy dissipation. We can thus reduce the energy injected in the quantum reservoir at each time step by reducing the number of cluster couplings (or their strength), by reducing the magnitude of the encoding values ($\alpha_{i}$ and $\beta_{i}$, $i = 1, \dots, N$) or by decreasing the BS transmissivity. Also, the value of $\lambda(\left(\mathcal{G},\vec{\alpha}, T \right)$ increases with the number of modes, $N$, so the hyperparameter values may need to be adjusted when scaling the size of the system.

\section{Static XOR} \label{app-XOR}
In this section, we elaborate further on the static XOR task. Concretely, an analytical explanation of why initial coherent states are needed for the protocol to be successful. We first consider the case in which the initial state is a 2-mode vacuum state, $\ket{\psi_{0}} = \ket{0} \otimes \ket{0}$, with covariance matrix $\Gamma_{0} = I_{4}$ and first-order moments equal to zero. We consider the mode dependent input encoding $\hat{p}'_{i}(s_{i}) = \hat{p}_{i} + \alpha s_{i} \hat{x}_{i}$ ($i = 1, 2$), with $\alpha$ being an arbitrary parameter equal among modes. From Eq. \eqref{2nd-moments-eq} we can write the second-order moments in this scenario, yielding
\begin{equation} \label{2nd-2modes-vaccum}
    O_{\text{2nd}}^{(\text{vac})}(s_{1},s_{2}) = \left(
    \begin{array}{cc}
        (1-T)\alpha^{2}s_{1}^{2} + 1 & 0 \\
        0 & (1-T)\alpha^{2}s_{2}^{2} + 1
    \end{array}
    \right) \ 
\end{equation}
We can see how there are no mode correlations, $\langle \hat{x}_{1} \hat{x}_{2} \rangle = 0$, and there are no terms $\propto s_{1}s_{2}$. Those are the reasons this reservoir cannot implement the XOR gate successfully. On the other hand, we consider a two-mode coherent state, $\ket{\psi_{0}} = \ket{\alpha_{x}=1} \otimes \ket{\alpha_{x}=1}$, with the same covariance matrix but first-order moment vector equal to $\mathbf{R}_{0} = (1, 0, 1, 0)^{\top}$. Then, the resulting second-order moment matrix takes the form
\begin{equation}
    O_{\text{2nd}}^{(\text{coh})}(s_{1},s_{2}) = \left(
    \begin{array}{cc}
        2(1-T)\alpha^{2}s_{1}^{2} + 1 & (1-T) \alpha^{2} s_{1} s_{2} \\
        (1-T) \alpha^{2} s_{1} s_{2} & 2(1-T)\alpha^{2}s_{2}^{2} + 1
    \end{array}
    \right) \ ,
\end{equation}
where the crossed term $s_{1} s_{2}$ is found in the quadrature correlation.
\section{NARMA$d$ task} \label{app-narma}
The NARMA$d$ task, where $d$ is a parameter determining the delay, requires approximating, at each time step, the target nonlinear function
\begin{equation} \label{narmad-eq}
    \bar{y}_{k}^{(d)} = \alpha \bar{y}_{k-1} + \beta \bar{y}_{k-1} \sum_{i=1}^{d} \bar{y}_{k-i} + \gamma u_{k-1} u_{k-d} + \delta \ ,
\end{equation}
where $u_{k} = \mu + \nu s_{k}$. The default set of constant parameters is set to $(\alpha, \beta, \gamma, \delta,\mu, \nu) = (0.3, 0.05, 1.5, 0.1, 0, 0.2)$ and every input value $s_{k}$ belongs to a uniform random distribution between -1 and 1. It was already shown that, for this set of parameters, any reservoir can reproduce the target function \eqref{narmad-eq} with low error as long as it can reach high linear memories up to delay $d$, $s_{k-d}$, and a quadratic memory of crossed terms up to the term $s_{k-1} s_{k-d}$ \cite{Kubota2021}.

\section{MNIST handwritten digit dataset} \label{app-MNIST}
In this section, we explain the details of the image classification task using the MNIST dataset. We start by discussing the data preprocessing of the images. Every image is first normalized, so every pixel takes values from zero to one. They are then compressed from 28x28 pixels to 14x14 pixels using the Zoning2 technique \cite{Antonik2020}. Each compressed pixel is the sum of 4 adjacent pixels (in a 2x2 window) of the original image. As the compressed images are also normalized, the full compression can be understood as taking the mean of 2x2 pixel windows of the original image. So, if $\mathcal{I}_{k}$ is a 28x28 matrix containing the pixels of the $k$-th normalized image in the dataset, then the compressed 14x14 image matrix, $\mathcal{I}_{k}^{(\text{comp.})}$, can be written as
\begin{equation}
\left[\mathcal{I}_{k}^{(\text{comp.})}\right]_{ij} = \frac{1}{4}\sum_{m,n=0}^{1}\left[ \mathcal{I}_{k} \right]_{2i+m-1,2j+n-1} \ ,
\end{equation}
for $i, j = 1, 2, \dots, 14$. Regarding the quantum reservoir, its initial state is set as a product of coherent states, $\ket{\Psi_{0}} = \bigotimes_{i=1}^{14} \ket{\alpha_{x}=1}$, as in the static XOR task. This is done to ensure the presence of crossed terms of the inputs from the same column. For the training protocol, we applied a softmax regression (Sect. \ref{softmax-app}) with 10 classes, one for each digit. For the optimization, we used an Adam Stochastic Gradient Descent algorithm \cite{kingma2017}, employing a mini-batch size of 32. The algorithm was implemented using the Tensorflow and Keras Python packages \cite{tensorflow2015-whitepaper,keras2015}. For every cluster topology, the reservoir was trained during 50 epochs, and the set of weights that achieved the highest accuracy on the test set was saved.
\end{document}